\documentclass[preprintnumbers,showkeys,superscriptaddress,showpacs,
nofootinbib,byrevtex,fleqn,prd,notitlepage]{revtex4-2}
\usepackage{amsmath,amsfonts,amssymb,amscd,amsxtra,amsthm}
\usepackage{graphicx}
\usepackage{bm,bbold}
\usepackage{multirow}
\usepackage{hyperref}
\usepackage[fleqn]{nccmath}
\usepackage{empheq}
\usepackage{cleveref}
\crefname{equation}{}{}
\input{epsf}
\newcommand{\be}{\begin{eqnarray}}
\newcommand{\ee}{\end{eqnarray}}
\newcommand{\ba}{\begin{array}}
\newcommand{\ea}{\end{array}}

\newcommand{\half}{{\textstyle{\frac{1}{2}}}}

\newcommand{\Slash}[1]{\ooalign{\hfil/\hfil\crcr$#1$}}

\newcommand{\gA}{g_{A}^{(3)}}
\usepackage[normalem]{ulem} 
\usepackage[dvipsnames]{xcolor} 
\renewcommand\sout{\bgroup \color{red} \ULdepth=-.5ex \ULset}

\begin{document}
\preprint{INHA-NTG-02/2022}
\title{Isovector helicity quark quasi-distributions inside a large-Nc nucleon\footnote{This paper is dedicated to Maxim Vladimirovich ($\dagger$August, 2021), who was my teacher, collaborator and sincere friend.}}
%
\author{Hyeon-Dong Son}
\email[ E-mail: ]{hdson@korea.ac.kr}
\affiliation{Center for Extreme Nuclear Matters (CENuM), 
Korea University, Republic of Korea}
\affiliation{Department of Physics, Inha University, Incheon 22212, Republic of Korea}
\affiliation{Ruhr-Universit\"at Bochum, Fakult\"at f\"ur Physik und Astronomie,
Institut f\"ur Theoretische Physik II, D-44780 Bochum, Germany
}

\begin{abstract}
In this letter, we report the results for the isovector polarized quark quasi- distribution functions $\Delta u(x,v) - \Delta d(x,v)$ in the large-$N_c$ limit, calculated within the chiral quark-soliton model. 
It is shown that the polarized quark quasi-distributions present good convergence to the light-cone PDFs in the limit of the nucleon boost momentum $P_N \to \infty$ (or the velocity $v \to 1$), compared to the case of the unpolarized isosinglet distributions.
\end{abstract}

\maketitle

\section{\normalsize \bf Introduction}
It is an intriguing and essential question to ask, how much portions of the nucleon spin are carried by each of its internal degrees of freedom; intrinsic spins of the valence and sea quarks and the gluons and their orbital angular momentum. Among them, the intrinsic spin content of the quarks and antiquarks can be accessed by the twist-2 longitudinally polarized quark distribution functions as their leading Mellin moments. In particular, in the isovector channel for the light quarks, we display the following definition of the corresponding parton distribution function (PDF), expressed as a Fourier transform of the QCD matrix element:
\begin{align}\label{eq:helicity_def}
    \Delta u(x) - \Delta d(x) = \int dz^- e^{i x z^- P^+}   \langle P | \bar{\psi}(0) W[0,z] \gamma^+\gamma_5 \tau^3 \psi(z) | P\rangle.
\end{align}
In the above, $W[0,z]$ is the gauge connection and $z$ is a light-like 4-vector in the coordinate space, $z^\pm=z^0\pm z^3$. $\psi$ is the light-quark field and $\tau^3$ picks up the isovector structure for the quarks. $x \in [0,1]$ is the quark momentum fraction to the nucleon momentum $P$ and the matrix element implicitly depends on the renormalization scale $\mu$. The quantity corresponds to the probability difference to find longitudinally polarized $u$ and $d$ quarks with a certain momentum fraction $x$ inside a nucleon, which is often called the flavor asymmetry. The antiquark distribution is related to Eq. \cref{eq:helicity_def} by the property $\Delta \bar u(x)- \Delta \bar d(x) =\Delta u(-x) - \Delta d(-x)$.

Especially the light-flavor asymmetry for the antiquarks has been a critical issue over decades. Until early 90's it was speculated that there is no difference for the longitudinally polarized anti- $u$ and $d$ quarks, so that $\Delta \bar u(x) - \Delta \bar d(x)$=0. Although there was no concrete proof for this conjecture, it has been often postulated and widely accepted; for instance, see Refs. \cite{Gluck:1995yq,Gluck:1995yr}.
On the other hand, a series of studies appeared, which predicts a significant amount of the antiquark flavor asymmetry \cite{Diakonov:1996sr,Diakonov:1997vc}, based on a large $N_c$ effective model of QCD at low energy. In Refs. \cite{Dressler:1998zi,Dressler:1999zg}, the authors pointed out that the DIS is not sensitive to the antiquark asymmetry, whereas the Drell-Yann process provides  a reasonable probe \cite{Dressler:1999zv,Kumano:1999bt}. 
The following analyses addressed the non-zero polarized antiquark asymmetry 
\cite{Gluck:2000dy,deFlorian:2009vb,Nocera:2014gqa}, utilizing deep inelastic scattering (DIS), semi-inclusive DIS, and Drell-Yann processes. 
It is known that the single-spin asymmetry for $W^\pm$ production in polarized proton-proton collisions is a tangible probe to the antiquark flavor asymmetry. 
 The idea is tested at the STAR collaboration at RHIC and reported the polarized antiquark asymmetry \cite{Adamczyk:2014xyw,Adare:2018csm,Adam:2018bam}. Also, there are upated PDFs reflecting these experimental improvements \cite{DeFlorian:2019xxt,Cocuzza:2022jye}.


Along with the experimental achievements, since Ref. \cite{Ji:2013dva} appeared, a great advance has been made in the lattice simulation of QCD to study the $x$ dependences of the PDFs. In the paper, X. Ji suggested a distinctive method for obtaining the objects on the light-cone from the lattice QCD, namely the large momentum effective theory (LAMET). The method applies to various quantities defined on the light-cone such as PDFs, distribution amplitudes, transverse-momentum dependent distributions and generalized parton distributions. 
For example, following Ref. \cite{Ji:2013dva}, one writes down the matrix element for so-called the quasi parton distribution function (quasi-PDF) corresponding to Eq. \eqref{eq:helicity_def}, with the spacelike separation $z=(0,0,0,z^3)$ of the quark bilinear operator and the nucleon carries a finite momentum $P_N$ in z-direction:
\begin{align}\label{eq:quasi-helicity_def}
    \Delta u(x,v) - \Delta d(x,v) = \int dz^3 e^{i x z^3 P^3}   \langle P | \bar{\psi}(0) W[0,z] \Gamma
    \gamma_5 \tau^3 \psi(z) | P\rangle.
\end{align}
The Dirac matrix $\Gamma$ can be either $\gamma^0$ or $\gamma^3$ as both approaches $\gamma^+$ in the limit  $P_N \to \infty$.
The essential observations are that,
the matrix element can be computed directly on the Euclidean lattice and
under the Lorentz boost $P_N \to \infty$, the matrix element approaches the corresponding light-cone one. 
The method has been utilized extensively to evaluate a wide class of 
the light-cone parton observables inside hadrons. 
For instance, the nucleon 
twist-2 PDFs are obtained at both the physical pion mass \cite{Chen:2018xof,Alexandrou:2018pbm,Lin:2018pvv,Alexandrou:2018eet,Liu:2018hxv,Alexandrou:2019lfo} and at the non-physical pion mass \cite{Liu:2018uuj,Fan:2020nzz,Alexandrou:2020qtt}. 
There exists a  alternative approach to obtain the light-cone PDFs on the lattice suggested by Radyushkin, {using the Ioffe-time distributions} \cite{Radyushkin:2017cyf}.
Also, there are numerous model studies on the quasi-PDFs and related topics \cite{Nam:2017gzm,Broniowski:2017wbr,Broniowski:2017gfp,Braun:2018brg,Bhattacharya:2018zxi,Bhattacharya:2019cme}.
We refer, for interested readers, to a review \cite{Cichy:2018mum} and a community report \cite{Constantinou:2020hdm}.

While the global analyses utilizing experimental results and the lattice simulations of QCD in the LaMET framework are competing and helping each other for a better understanding of the nucleon PDFs, a complementary model
study would be very useful as it is expected to provide a detailed description or picture how such nonperturbative nature is emergent from QCD in terms of effective model languages. 
One of such study is made for the twist-2 nucleon quasi-PDFs in Ref. \cite{Son:2019ghf}, which pointed out that the similar concept as quasi-PDF was already used to obtain the light-cone quark distribution functions in the nucleon from the chiral quark-soliton model ($\chi$QSM) \cite{Diakonov:1996sr,Diakonov:1997vc}. 
In the paper, the authors showed that one can define the quark quasi-distribution functions within the $\chi$QSM in the same manner as given in Ref. \cite{Ji:2013dva}. Also, they examined the properties of the twist-2 nucleon quasi-PDFs such as sum-rules as their Mellin moments and the nucleon momentum evolutions. 
In the present work, we provide the numerical results for the isovector longitudinally polarized quark (quasi-)distributions and discuss their properties in detail. It is an interesting question that how the quality of the nucleon momentum evolution differs for isosinglet and isovector quasi-PDFs to their light-cone correspondences. Thus we compare the numerical results for the isovector polarized quasi-PDFs with the previously presented isoscalar unpolarized ones in Ref. \cite{Son:2019ghf} in detail, to provide insight how much nucleon momentum is required for the LaMET framework. 

Let us shortly outline the present work. Firstly in Section \ref{sc:chqsm}, we briefly describe the main theoretical tool of the present study: the $\chi$QSM. In the following section, Section \ref{sc:IVP}, the model expressions of the helicity quasi-PDFs are provided and their properties are explored. In Section \ref{sc:result}, the numerical results are shown and discussed, focusing on the nucleon momentum evolution $P_N \to \infty$ of the quasi-PDFs. After that, we come to the conclusions and summary.

\section{\normalsize \bf Chiral quark-soliton model}\label{sc:chqsm}
The nucleon can be recognized as a chiral soliton in the large $N_c$ limit \cite{Witten:1979kh}, 
where the quarks inside the nucleon are project to a common static mean field as a result of their nonlinear interactions due to the dynamically broken chiral symmetry of the strong interaction. 
Meanwhile the detailed form of the quark effective interactions can be traced from QCD via the
instanton approach \cite{Diakonov:1983hh,Diakonov:1985eg}, relying on the small instanton packing fraction $\pi^2(\bar \rho / \bar R)^4/2\sim 1/16$, in general one can write down the following partition function respecting the chiral symmetry and its dynamical breakdown, in the Euclidean space-time: 
\begin{align}\label{eq:partition}
    Z[\psi^\dagger, \psi, \pi^a] = \int \mathcal{D}\pi^a \mathcal{D}\psi \mathcal{D}\psi^\dagger
    \exp \left[\int d^4x \psi^\dagger (i\Slash{\partial} + i M \exp(i \pi^a \tau^a \gamma^5)) 
    \psi \right].
\end{align}
Here, $M$ is the dynamical quark-mass resulting from the spontaenous chiral symmetry breaking. Note that the instanton picture of QCD provides a detailed form of the quark-momentum dependence $M(k)$, where its numerical value at zero quark momentum $M(k=0)=345~$MeV is determined by solving the gap equation. In this work, we simply ignore the quark momentum dependence and treat the dynamical quark mass as a constant quantity $M=M(k=0)$. Also, we only consider the flavor $SU(2)$ symmetry case and assume the chiral limit $m_\pi=0$,
 where the current quark-mass is neglected.

Integrating the fermion fields in Eq. \cref{eq:partition}, one obtains the effective action $S_{\mathrm{eff}}$ in the following form,
\begin{align}\label{eq:action}
 S_{\mathrm{eff}} = - N_c \mathrm{Tr} \ln \left[i\Slash{\partial} +i  M U^{\gamma_5} \right].
\end{align}
Note that a notation for the chiral field is used for convenience in \eqref{eq:action},
\begin{align}
    U^{\gamma_5}(x) = \frac{1+\gamma_5}{2}U(x)+\frac{1-\gamma_5}{2}U(x)^\dagger, \quad U(x) \equiv \exp(i \pi^a(x) \tau^a).
\end{align}
For the pion mean field, we introduce the hedgehog ansatz,
\begin{align}\label{eq:hedgehog}
   \pi^a(x) \tau^a \to \hat n ^a P(r) \tau^a.
\end{align}
In such configuration, the symmetries of the spatial and isospin rotations are not conserved individually. Instead, so-called the grand-spin $K\equiv J+T$ becomes a good quantum number of the Dirac hamiltonian $h(U)$ of the theory in a way that
the spatial rotation is compensated by the simultaneous isospin rotation (hedgehog symmetry). There exists as well the parity quantum number $P=(-1)^K$ (natural parity) and $P=(-1)^{K+1}$ (unnatural parity) which commutes with the Dirac hamiltonian. 
The static Dirac equation should be diagonalized
\begin{align}
    h(U)\Phi_n(\vec x) = E_n \Phi_n(\vec x),
\end{align}
to find the energy eigensolutions of the system.
Among the Dirac spectrum, there exists a distinctive level with the quantum number $K^P = 0^+$ which emerges from the upper Dirac continuum as the chiral mean field is formed. The $N_c$ quarks occupying the level generate the common mean field $U$ that polarizes the Dirac vacuum spectrum. The polarized Dirac sea, then, affects again the spectrum of the levels. As a result of the self-consistent feedback process of the level and the Dirac sea, the level becomes bound and one finds the minimum of the energy functional, which can be obtained from the leading exponent of the nucleon-nucleon correlation function in large Euclidean time. 
The corresponding classical soliton energy at the saddle point $U_{cl}$ has the following expression,
\begin{align}\label{eq:MN}
    M_{cl}  = N_c E_{\mathrm{level}} (U=U_{cl})
    + \sum_{n<0} E_{n}(U=U_{cl}) - \sum_{n<0} E_{n}(U=1).
\end{align}
The coherent summation of the negative levels including the last term, namely the vacuum subtraction, corresponds to the accumulated energy of the Dirac continuum due to the polarized vacuum. One should note that the \textit{continuum energy} possess the logarithmic divergence and thus 
introduce a regularization scheme. 
Note that the classical soliton does not carry the correct quantum number of the nucleon. The realistic quantum numbers of the nucleon such as the isospin and the total angular momentum are acquired after quantizing the soliton with respect to the rotational zero-modes. The detailed procedure can be found in Refs. \cite{Diakonov:1987ty,Christov:1995vm}.

\section{\normalsize \bf Helicity quark quasi-distributions in the large-$N_c$ limit}
\label{sc:IVP}
In this section, we use the result for the helicity quasi-PDFs in the$\chi$QSM \cite{Son:2019ghf} and discuss their properties. The detailed calculations and discussions can be found in Refs. \cite{Diakonov:1996sr,Diakonov:1997vc,Son:2019ghf}.

For the quasi-PDFs, the momentum fraction 
$x=k_3/P_N$ can have any value in $x\in (-\infty, \infty)$ in Eq. \cref{eq:quasi-helicity_def} as the nucleon is off the light-cone.
For simplicity, we use the nucleon velocity $v$ in the third direction made, which can be easily converted to the nucleon momentum $P_N = M_N v / \sqrt{1-v^2}$. 
In the model framework, the isovector longitudinally polarized quark quasi-distribution has the following form \cite{Son:2019ghf}:
\begin{align}\label{eq:IVP_model}
     \Delta u(x,v) - \Delta d(x,v) &= - \frac{1}{3} (2T^3)  N_c M_N v  
     \sum_{n,occ} \int \frac{d k^3 d^2k_\perp}{(2\pi)^3}
     \delta(k^3 + v E_n - v M_N x) \cr
     & \qquad \qquad \qquad \qquad \qquad \qquad
      \bigg[ \Phi^\dagger_n(\vec k) (1 + v \gamma^0 \gamma^3) \gamma^0  
    \Gamma \tau^3 \gamma^5 \Phi_n(\vec k)\bigg],
\end{align}
where $\Gamma$ is for $\Gamma=\gamma^0$ or $\Gamma = \gamma^3$, depending on the choice of the Dirac matrix in Eq. \cref{eq:quasi-helicity_def} and $T^3$ is the third component of the nucleon isospin. We consider only $T^3=1/2$, the proton case.
Some comments would be useful: firstly, the overall factor $N_c M_N$ reflects the $N_c$ counting of the (quasi-)PDF $\Delta u - \Delta d \sim N_c^2$, as $M_N \sim N_c$ in the large $N_c$ limit. Secondly, the infinite summation over the occupied Dirac continuum should be subtracted by its vacuum equivalent (vacuum subtraction). The remaining piece diverges logarithmically and the treatment of such divergence should be consistent with the minimizing procedure of the action, as described in the previous section.
Finally, we check that the result approaches smoothly to the light-cone PDF in the limit $v \to 1$, or $P_N \to \infty$. 

Now let us move on to the leading Mellin moment of the isovector polarized distribution, or the Bjorken spin sum-rule.
By integrating Eq. \cref{eq:IVP_model} over $x$ from $-\infty$ to $\infty$, we obtain the following result \cite{Son:2019ghf},
\begin{align}
    \label{eq:spin_sum}
    \int^{+\infty}_{-\infty} dx (\Delta u(x,v) - \Delta d(x,v) )=
    \bigg\{ \begin{split}
        v g_A^{(3)}, \Gamma=\gamma^0 \\
         g_A^{(3)}, \Gamma=\gamma^3
    \end{split} \bigg\},
\end{align}
where the isovector axial charge has the following expression in the model,
\begin{align}\label{eq:ga_model}
    g_A^{(3)}= - \frac{1}{3}  N_c  
    \sum_{n,occ} \int \frac{d k^3 d^2k_\perp}{(2\pi)^3}
    \bigg[ \Phi^\dagger_n(\vec k)  \gamma^0 \gamma^3 \gamma^5 
    \tau^3\Phi_n(\vec k)\bigg]. 
\end{align}
As already discussed in Ref. \cite{Son:2019ghf}, the sum-rule \cref{eq:spin_sum} is a natural result, considering the local operator
corresponding to the l.h.s and related spin matrix of the nucleon. Notice that the nucleon velocity factor $v$ appears in the case of $\Gamma=\gamma^0$. From this, we predic that the isovector polarized quark quasi-distribution defined with the Dirac matrix $\Gamma=\gamma^3$ exhibits better convergence to the light-cone one under the Lorentz boost $P_N \to \infty$. 

\section{\normalsize \bf Numerical results and discussions}\label{sc:result}
In this work, to avoid complications in the numerical computation, we adopt an approximation method 
which is called the interpolation formula. In the approximation scheme, one takes $pM(U-1)/(p^2+M^2)\ll 1$ as a small expansion parameter, in which three different limits are valid: 1. the pion momentum is small $p/M\ll 1$, where p denotes the characteristic momentum of the pion, 2. the pion momentum is large: $p/M \gg 1$, and 3. the pion field is small $U \approx 1$. 
{Note that, in the case of the light-cone PDFs, numerical computation of the vacuum contributions can be avoided by selecting an appropriate negative- or positive-level summation scheme due to a simple kinematic consideration \cite{Diakonov:1996sr,Diakonov:1997vc}. However, in the case of the quasi-PDFs, we observe that such '\textit{vacuum-safe}' summation depends also on the momentum of the nucleon and one cannot avoid the computation of the vacuum contribution for particular $x$ region. Normally, the full numerical calculation proposed in \cite{Diakonov:1997vc} becomes quite involved when one has to explicitly include the vacuum contribution. By using the interpolation formula, one can reduce a great amount of the computation time.}
Also, we use the following parametrized ansatz for the pion mean field in Eq. \cref{eq:hedgehog}:
\begin{align} \label{eq:arctan}
    P(r) = 2\mathrm{ArcTan}\left(\frac{r_0^2}{r^2}\right),
\end{align}
where $r_0 \approx 0.98/M$ is determined by the variational approach \cite{Diakonov:1987ty}. While the deviation for various nucleon observables with such ansatz from using the self-consistent saddle-point solution \cref{eq:MN}  is known to be within $\sim 10 \%$ \cite{Diakonov:1987ty,Weiss:1997rt}, the complexity of the numerical computation is significantly reduced \cite{Diakonov:1987ty,Diakonov:1997vc}. 
Also, using the ansatz makes one to explorer the properties of the physical quantities analytically. For instance, if one only takes into account the limit $p/M \ll 1$, the familiar chiral lagrangian is obtained, with the low energy constants are expressed in terms of the quark loop integrals.

As described in Ref. \cite{Son:2019ghf} in detail, Eq. \cref{eq:IVP_model} possess the logarithmic divergences
due to the polarized vacuum or the continuum contribution. We tame this divergence by adopting the Pauli-Villars(PV) regularization method with the single-subtraction where the PV mass $M_{\mathrm{PV}}\approx 560~$MeV is fixed from the mesonic sector of the model to reproduce the empirical pion decay constant,
\begin{align}
    F_\pi^2 = \frac{N_c M^2}{4\pi^2} \ln \left( {M_{\mathrm{PV}}^2/M^2} \right)=(93~\mathrm{MeV})^2.
\end{align}

\begin{figure}[htbp]
    \centering
    \includegraphics[width=7.5cm]{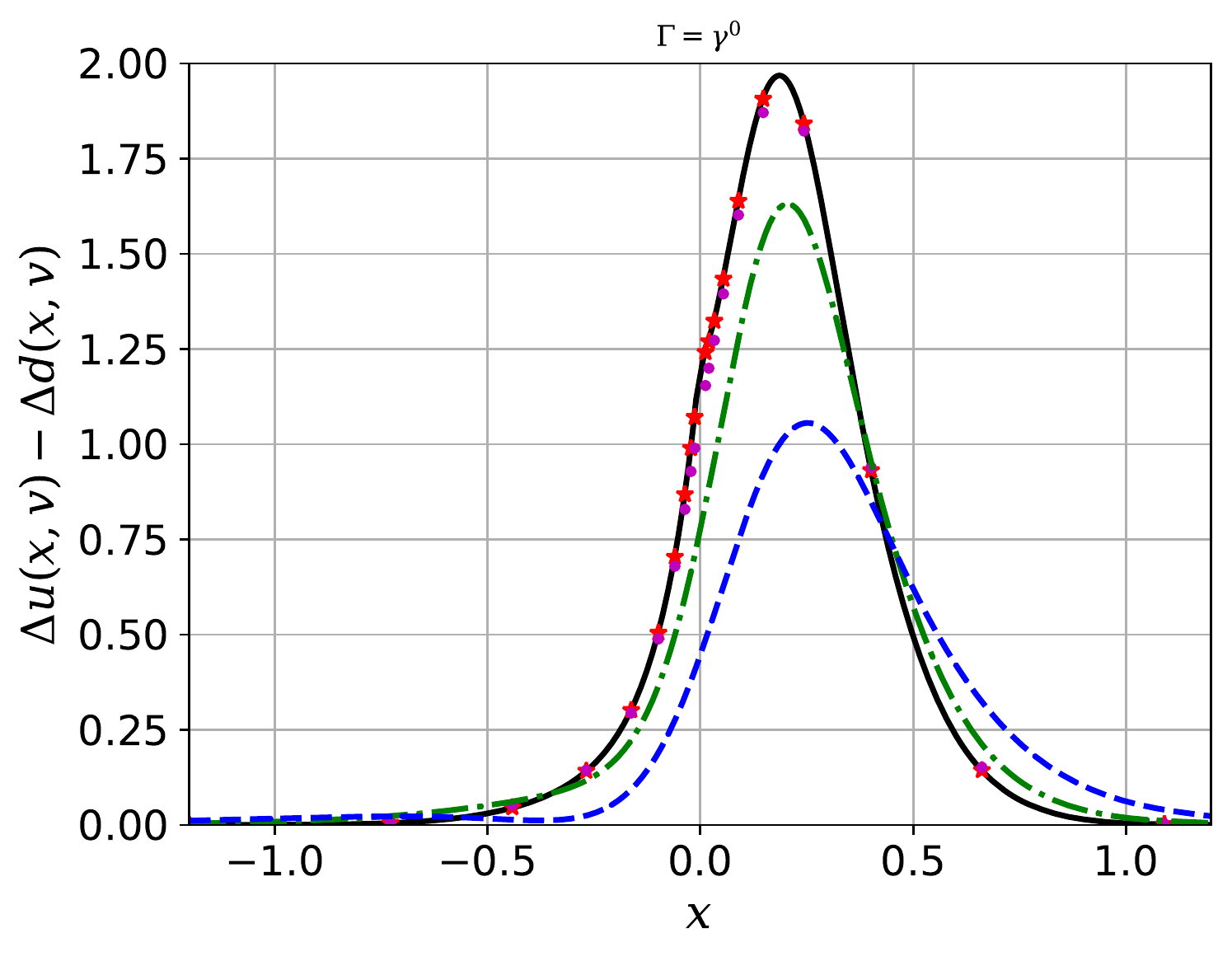}
    \includegraphics[width=7.5cm]{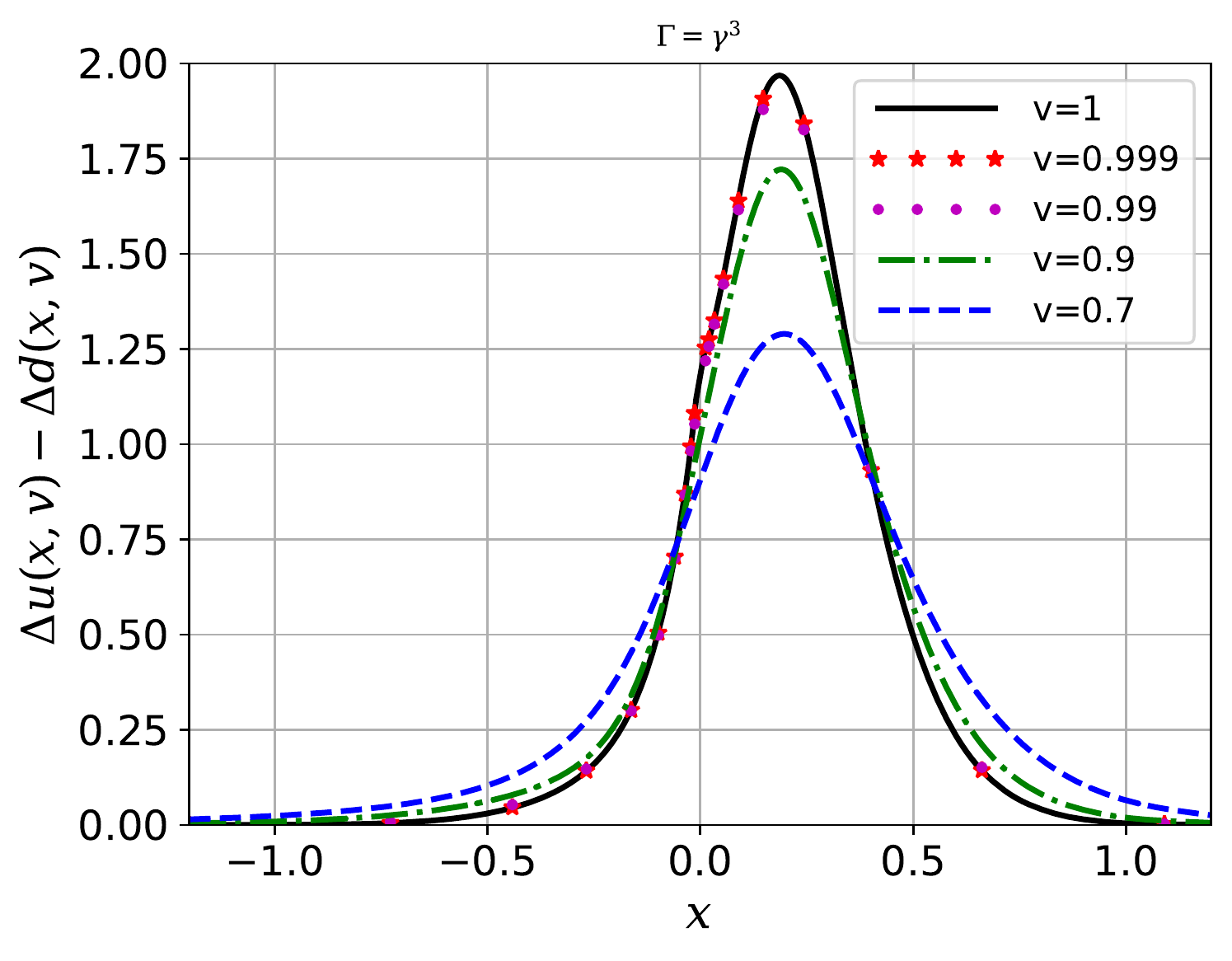}
    \includegraphics[width=12cm]{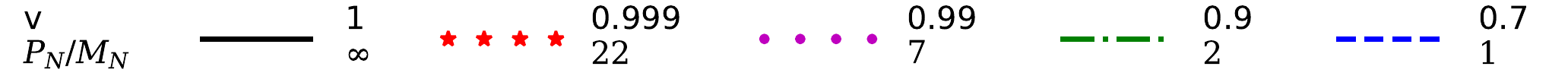}
    \caption{Isovector polarized quasi-PDFs for quarks and antiquarks in the nucleon. $\Gamma=\gamma^0$(left) and $\Gamma=\gamma^0$(right)}
    \label{fig1}
\end{figure}
\begin{figure}[htbp]
    \centering
    \includegraphics[width=9cm]{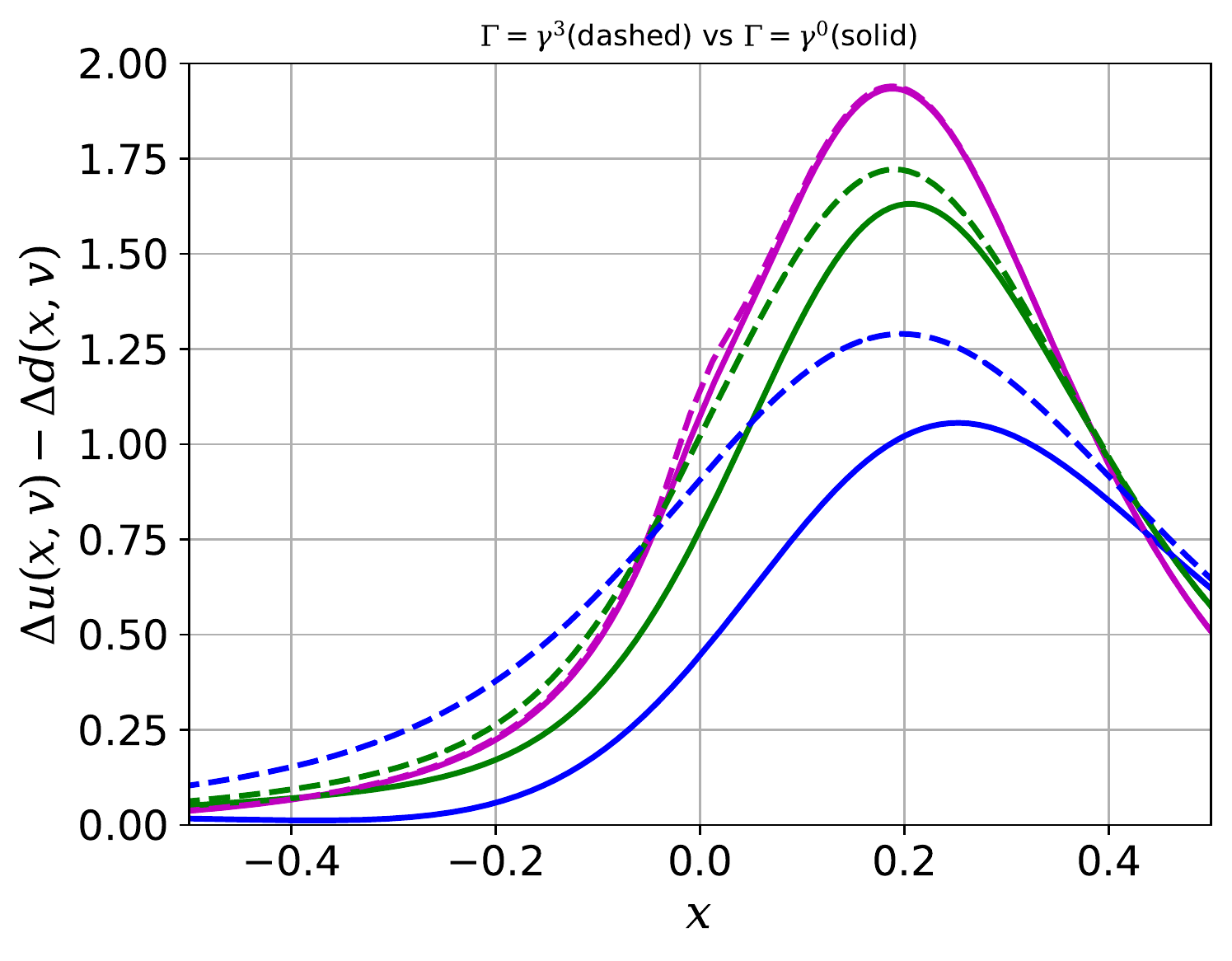}
    \caption{Isovector polarized quasi-PDFs for quarks and antiquarks in the nucleon with $\Gamma=\gamma^0$(left) and $\Gamma=\gamma^0$(right). Same color scheme is used as Fig. \ref{fig1}; Purple ($v=0.99$), Green ($v=0.9$), and Blue ($v=0.7$). }
    \label{fig2}
\end{figure}

In Fig. \ref{fig1}, we display the numerical results for the quark quasi-distributions with the Dirac matrix $\Gamma=\gamma^0$(left) and $\Gamma=\gamma^0$(right), respectively, for selected values of the nucleon boost velocities. One observes that the Dirac structure $\Gamma=\gamma^3$ provides a better convergence to the light-cone distribution function, which can also be predicted from the result of the Bjorken spin sum-rule \cref{eq:spin_sum}. A magnified view at smaller x region ($|x| \lesssim  0.5$) is given in Fig. \ref{fig2}, for a comparison of $\Gamma=\gamma^0$(solid) and $\Gamma=\gamma^3$(dashed).
It is shown that at rather small nucleon momentum $P_N \approx 1.4 M_N$ ($v=0.7$, blue curves in Fig. \ref{fig2}) the difference between the $\Gamma=\gamma^0$ and $\Gamma=\gamma^3$ distributions is quite noticeable. 
We remind that, in the lattice QCD, there exist technical limitations taking the boost to the infinite momentum frame due to, for instance, the signal-to-noise ratio and the excited states contamination \cite{Cichy:2018mum}. 
  Instead, the quasi-PDFs are computed at a finite nucleon momentum, typically  $P_N \sim 2$ GeV and the light-cone PDFs are obtained by using the the perturbative matching procedure.
In the renormalization process, the Dirac structure defining the quasi-PDFs may involve mixing with other operators \cite{Constantinou:2017sej}. {In particular, it was shown that in the case of the unpolarized distributions, 
$\gamma^3$ presents a mixing with the scalar operator (twist-3).
This peculiar behavior suggests one to use
the Dirac operator with $\gamma^0$ for the unpolarized distributions.} 
On the other hand, in the case of the polarized distributions, 
$\gamma^3$ should be used \cite{Constantinou:2017sej}.

\begin{figure}[htbp]
    \centering
    \includegraphics[width=7.5cm]{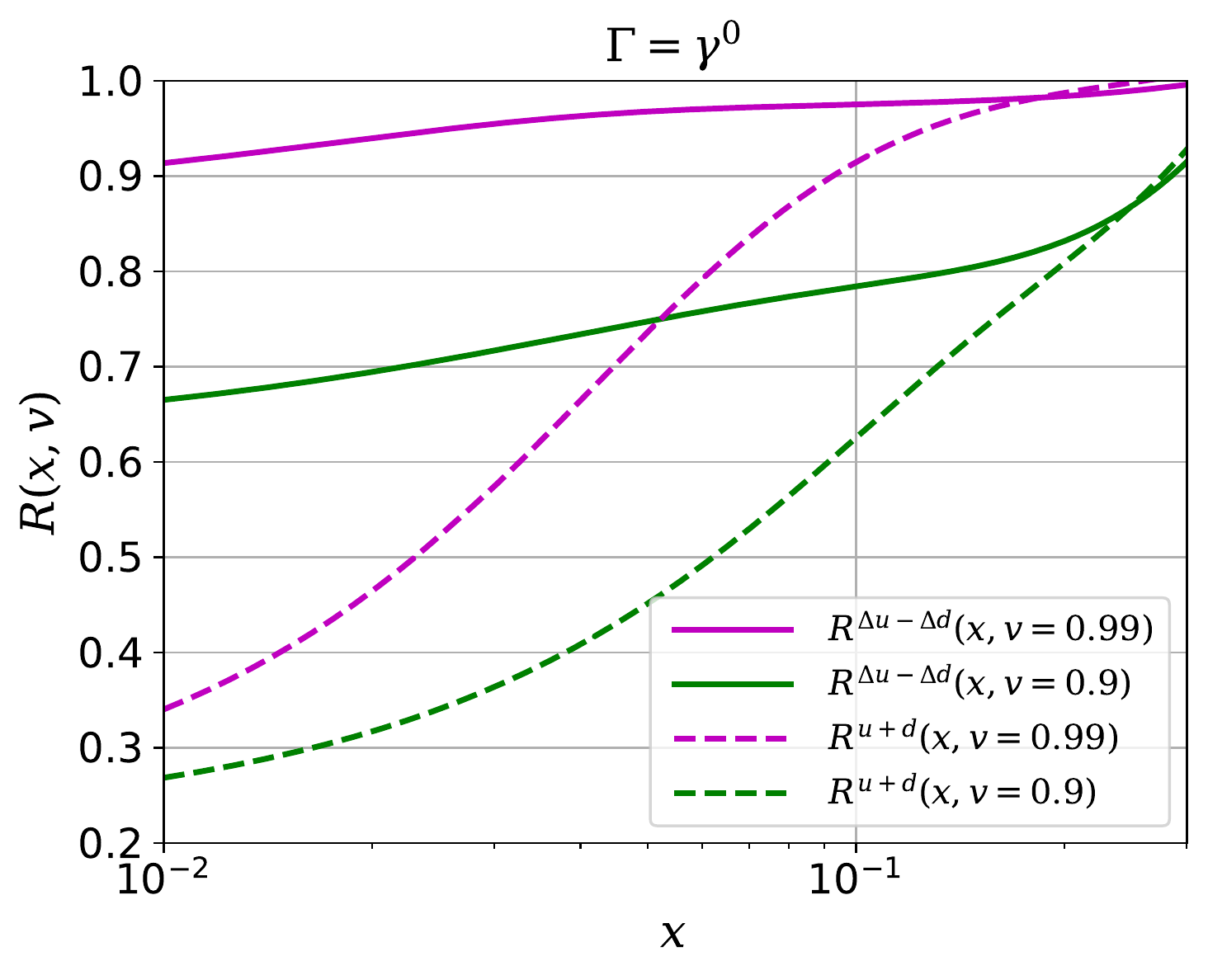}
    \includegraphics[width=7.5cm]{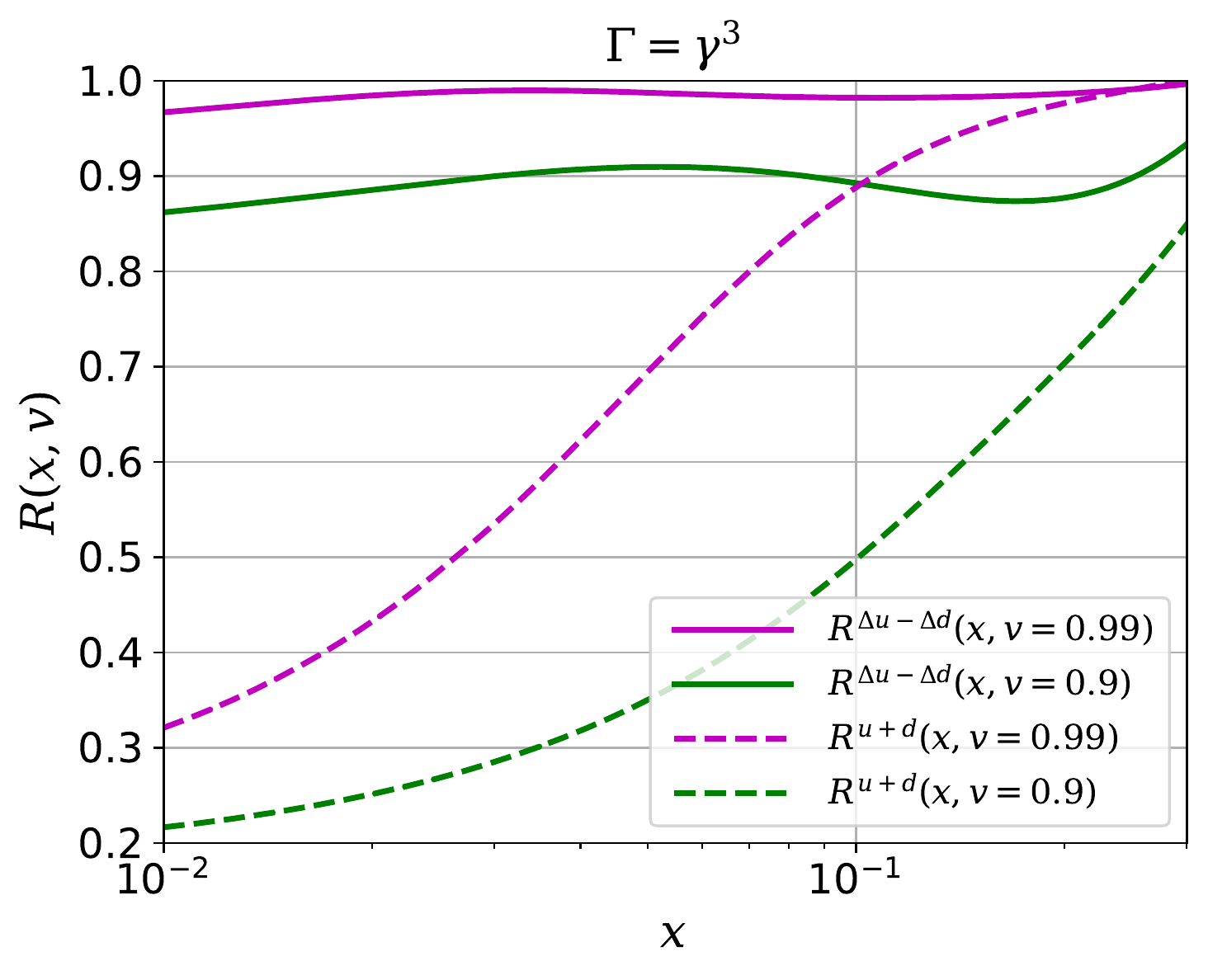}
    \caption{Ratios $q(x,v)/q(x,v=1)$, for isovector polarized (solid) 
     and isoscalar unpolarized (dashed, \cite{Son:2019ghf})  quasi distributions, for $v=0.99$ (magenta) and $v=0.9$ (green), and for $\Gamma=\gamma^0$ (left) and $\Gamma=\gamma^3$ (right). }
    \label{fig3}
\end{figure}

\begin{figure}[htbp]
    \centering
    \includegraphics[width=7.5cm]{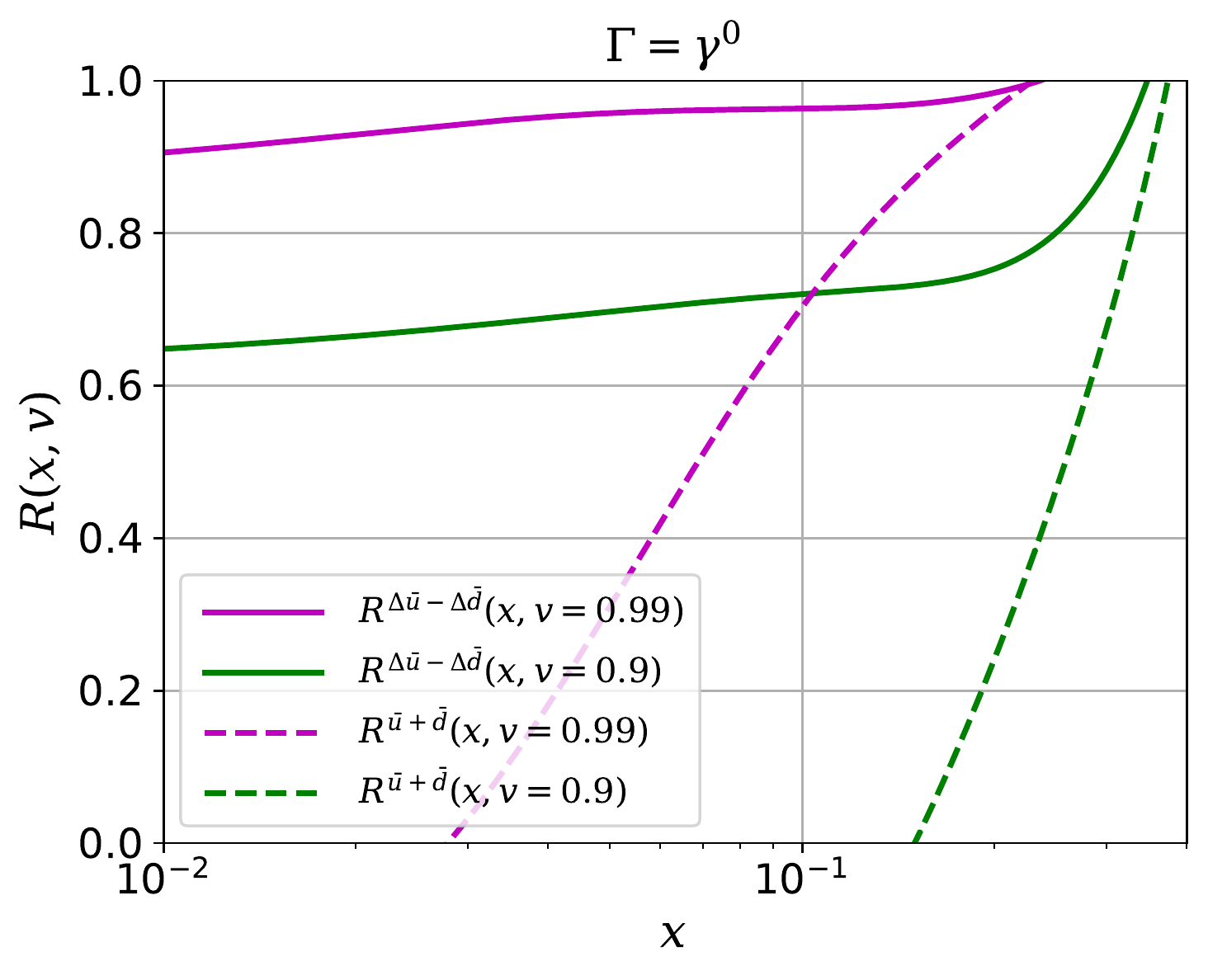}
    \includegraphics[width=7.5cm]{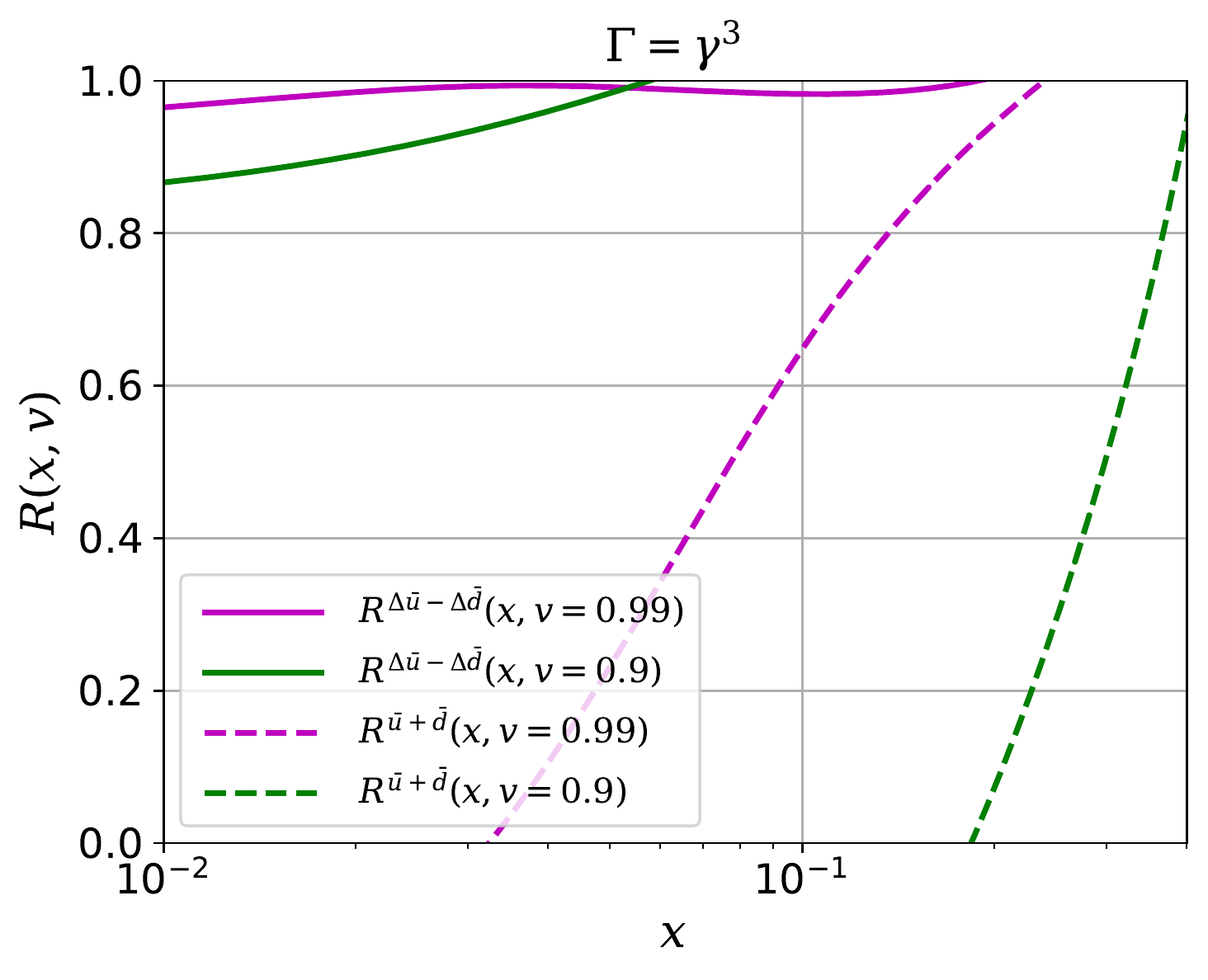}
    \caption{Ratios $\bar q(x,v)/ \bar q(x,v=1)$, for isovector polarized (solid) 
    and isoscalar unpolarized (dashed, \cite{Son:2019ghf}) quasi distributions, for $v=0.99$ (magenta) and $v=0.9$ (green), and for $\Gamma=\gamma^0$ (left) and $\Gamma=\gamma^3$ (right). }
    \label{fig4}
\end{figure}

Comparing the current numerical results with the isoscalar unpolarized quasi distributions presented in Ref. \cite{Son:2019ghf}, we find that the isovector polarized distributions approach faster the light-cone ones in the limit $v \to 1$, regardless of the Dirac matrices defining them. The origin of this behavior is the distinct characteristics of the matrix elements on the light-cone, $\bar q(x) = - q(-x)$ for the isoscalar unpolarized one and $\Delta \bar q(x) = \Delta q(-x)$ for the isovector polarized one. As the nucleon is off the light-cone, the quarks and antiquarks can have negative momentum, and thus the quark and antiquark quasi-distributions smear outside the cannonical support of $x \in [0,1]$. Such smearing influences significantly to the isoscalar unpolarized distributions $u(x,v)+d(x,v)$, leading sharp dips at small $|x|<0.1$ region. Note that the polarized Dirac sea governs this property, see Appendix \ref{app.1}, where the level and the continuum contributions are given separately for both quasi-PDFs.
To observe the tendency graphically, we define the ratio function $R^{q} (x,v)\equiv q(x,v)/q(x,v=1)$ for a generic quasi-PDF $q(x,v)$. In Fig. \ref{fig3}, $R^{u+d}(x,v)$ and $R^{\Delta u - \Delta d}(x,v)$ for are compared for the nucleon boost velocities $v=0.99$ (magenta) and $v=0.9$ (green). In Fig. \ref{fig4}, the same comparison is made for the $x<0$ region, or the antiquarks. As the isoscalar quasi-distributions become negative \cite{Son:2019ghf}, the differences of the isoscalar (solid) and the isovector (dashed) ratios are even larger compared with the plots in Fig. \ref{fig3}. Note that, as one can see in Fig. \ref{fig1}, the quasi-distributions at negative $x$ are positive even for the smallest nucleon boost momentum $P_N \approx M_N $ ($v=0.7$, blue dashed-curves).
 Overall, we conclude that the isovector polarized quasi-distributions (solid lines) have much better convergence to the light-cone PDFs. 
For discussions about the large-$x$ behavior of the quasi-PDFs in the large $N_c$ limit, we refer \cite{Son:2019ghf} and the references therein. 

\begin{figure}[htbp]
    \centering
    \includegraphics[width=6.5cm]{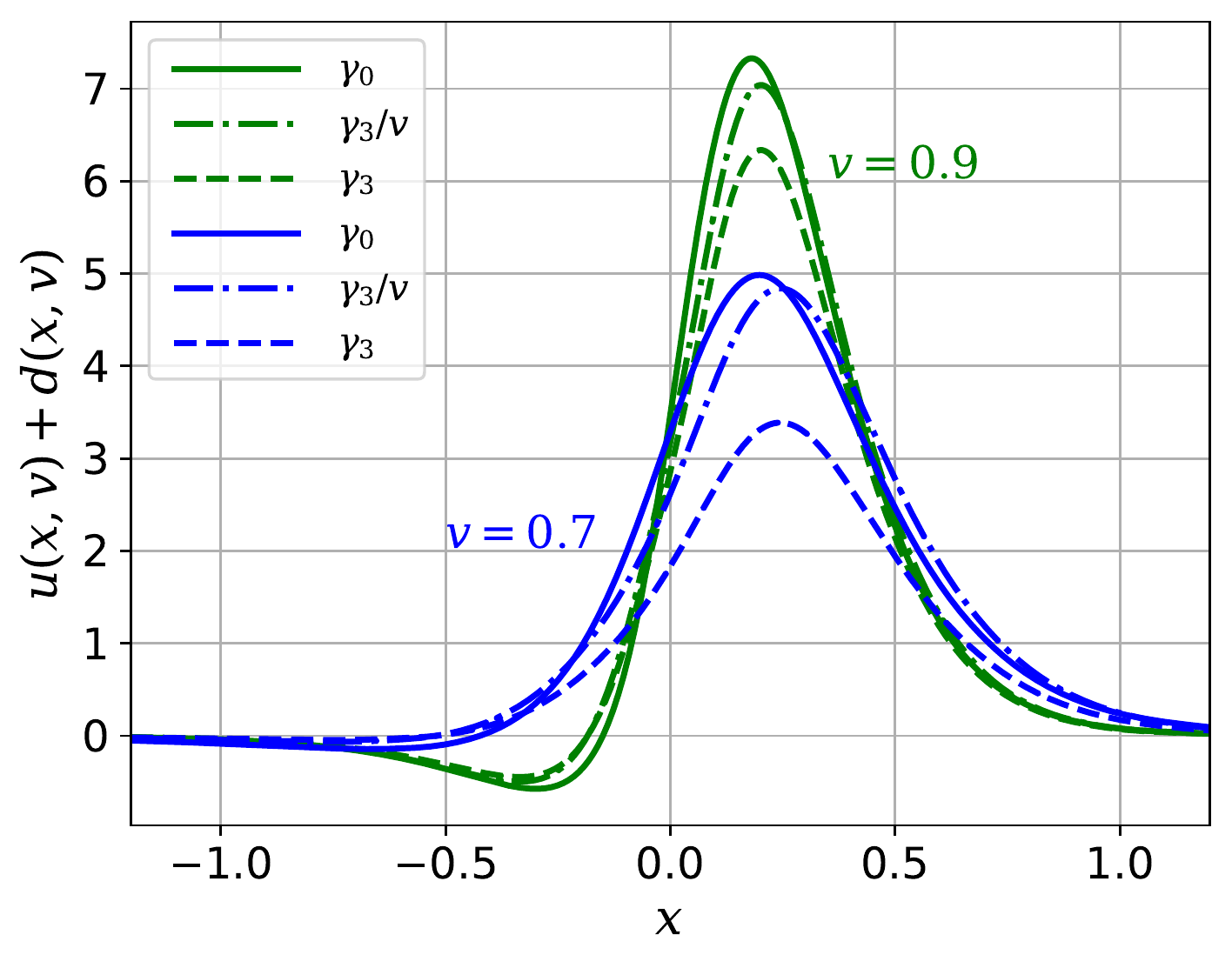}
    \includegraphics[width=6.8cm]{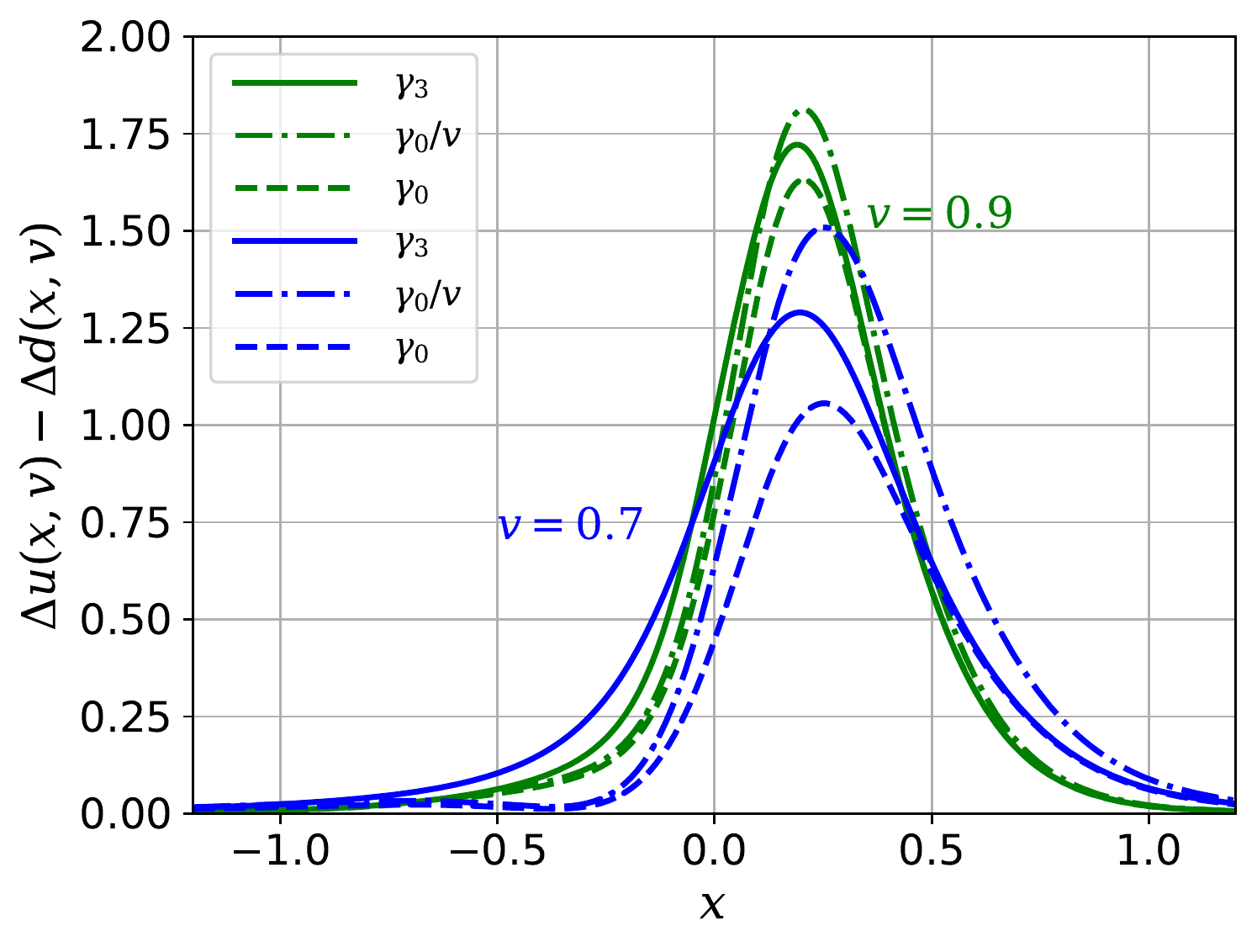}
    \caption{$u(x,v)+d(x,v)$s defined with 
    $\Gamma=(\gamma_0,\gamma_3/v,\gamma_3)$ are plotted on the left panel.
    On the right panel, $\Delta u(x,v) - \Delta (x,v)$s defined with 
    $\Gamma=(\gamma_3,\gamma_0/v,\gamma_0)$ are shown.}
    \label{fig5}
  \end{figure}

Note that the definitions of quasi-PDFs are arbitrary as far as they 
present correct limit to the light-cone PDFs. For example, one may redefine 
the quasi-PDFs with aditional factor of the nucleon boost velocities, 
for instance $\Gamma =\gamma^3 \to \Gamma = \tfrac{1}{v}\gamma^3$ 
for the isoscalar unpolarized distributions. We examine that the 'new' isoscalar unpolarized quasi-PDFs with $\Gamma = \tfrac{1}{v}\gamma^3$
 are comparable to those with $\Gamma =\gamma^0$, as shown 
 in the left panel of Fig. \ref{fig5}. 
We may consider the isovector polarized quasi-PDFs with $\Gamma = \tfrac{1}{v}\gamma^0$. In this case, the factor $1/v$ plays more significant role
as the quasi-PDFs with $\gamma^0$ and $\gamma^3$ are close to each other.
 We observe that the quasi-PDFs with $\Gamma = \tfrac{1}{v}\gamma^0$ are better than those with $\Gamma=\gamma_3$ considering the convergence to the light-cone PDFs.

In Ref  \cite{Bhattacharya:2018zxi}, it is mentioned that the second Mellin moment of the unpolarized quasi-PDF (momentum sum-rule) depend on 
the kinematic variable $\delta_0 \equiv 1/v$, through the 
nonconserved energy-momentum tensor(EMT) form-factor $\bar c$. 
Thus, in general, one cannot remove the nucleon momentum 
dependence of the Mellin moments of the quasi-PDFs by redefining 
them. However, the current model considers only the quark degrees of freedom and one observes that the momentum sum-rule is only satisfied by
 quarks and antiquarks. Accordingly, the quark EMT operator is conserved and 
 the form-factor $\bar c$ vanishes.

One should keep in mind that the current result only involves the leading order contribution in the large $N_c$ limit. Hence one obtains the following numerical value of the isovector axial charge,
\begin{align}\label{eq:ga}
    \gA \approx 0.9,
\end{align}
a much underestimated prediction compared with the phenominological value $\gA = 1.28$ \cite{ParticleDataGroup:2020ssz}.
It is known that the contribution from the next-order terms in the large $N_c$ expansion are significant to the isovector axial charge $\gA$ \cite{Christov:1995vm}. For instance, the $\chi$QSM predicts \cite{Blotz:1994wi},
\begin{align}\label{eq:ga_chqsm_full}
    \gA = 0.92 (N_c) + 0.38 (N_c^0)\;,
\end{align} 
 where the small difference of $\sim 1\%$ from the current result \cref{eq:ga} may come from the different regularization schemes, the use of different dynamical-quark mass, and the use of the ArcTan ansatz \cref{eq:arctan} for the pion mean field. Note as well that the naive quark model result for the axial charge also predicts rather large $1/N_c$ correction: $\gA(\mathrm{QUARK\; MODEL}) = N_c/3 + 2/3$. Thus, when the $1/N_c$ correction is added to the current result, the overall magnitude of the quasi-PDF should be increased as well, accordingly. 

The numerical results of the current study
may have a few sources of systematic uncertainties.
One of the source is that the interpolation formula and the 
arctan ansatz \cref{eq:arctan} are used for numerical computation.
Another point is that the dynamical quark mass $M$ may vary slightly 
according to the chocie of the instanton variables $\bar \rho$ and $\bar R$
 and their ratio, see for example \cite{Diakonov:1995qy,polyakov:2020cnc}. 
Finally, there could be regularization scheme dependences. 
We anticipate that their contributions can add up to order of $~10\%$
and would not change the major numerical observations 
at the current order of $1/N_c$ expansion in the large $N_c$ limit.  
We emphasize that the main observation of the present work is that the 
helicity quasi-PDFs have better convergence to the isoscalar unpolarized 
once.  We argue that this distinction stems from the nature of the 
quark distribution functions as explained earlier, 
and thus quite model independent. 


\section{\normalsize \bf Summary}
In this letter,
     we presented the numerical results on the isovector longitudinally polarized  quark quasi-distributions inside the nucleon, computed within the $\chi$QSM. In general, the results can be considered as the initial point of the renormalization scale evolution of QCD for the quark quasi-distributions where the scale is given by the inverse of the average instanton size $1/\bar \rho \approx 600 ~$MeV. We focused on their evolution under the boost of the nucleon momentum to the light-cone $P_N\to \infty$.
     In \cite{Son:2019ghf}, we presented the numerical result for the isoscalar unpolarized quasi distributions $u(x,v)+d(x,v)$. One of the major observations in the paper is that the unpolarized quasi-PDF approaches the light-cone one slowly, especially at smaller $x$ region, $x \lesssim 1/N_c$. 
     In the current study, we observed that the isovector polarized quasi-PDFs have better convergence to the light-cone ones, compared with the isoscalar unpolarized distributions.
     We provided a simple explanation of such behavior by comparing it to the isovector polarized quasi-PDF $\Delta u(x,v) - \Delta d(x,v)$, considering the relations for the quark and antiquark PDFs and their smearing out of the support $x\in [0,1]$, when the nucleon is off the light-cone.
     The Bjorken spin sum-rule, depending on the Dirac matrix defining the quasi-PDF predicts that $\Gamma=\gamma^3$ is a better choice than $\Gamma=\gamma^0$ in terms of the nucleon momentum evolution,
     unless the alternative definition $\Gamma=\gamma^0/v$ is used.



\section*{\normalsize \bf Acknowledgement}
HDS is grateful to Maxim V. Polyakov who initiated this series of studies. The author also thanks to Asli Tandogan, June-Young Kim and Hyun-Chul Kim for the fruitful discussions and comments. 
This work is supported by the DFG through the Sino-German CRC 110 “Symmetries and the Emergence of Structure in QCD” and Basic Science Research Program through the National Research Foundation of Korea funded by the Korean government (Ministry of Education, Science and Technology, MEST), Grant-No. 2018R1A5A1025563 and 2021R1A2C2093368.

\appendix
\section{Level and continuum contributions}\label{app.1}
\subsection{Level contributions}
It was mentioned in Section \ref{sc:chqsm} that there exists a
unique Dirac level that is bound by the self-consistent mean-field
and has the Grandspin and parity $K^P=0^+$. In this chapter, we provide the detailed form of the level contributions and their plots.
The notation is taken from Ref. \cite{Diakonov:1996sr}.
In the hedgehog ansatz, the Dirac wave function for the bound level only depends on the radial component in the spherical coordinate system and has the following form,
\begin{align}
    \Phi_{\mathrm{level}}(r) = \bigg(
        \begin{split}
            h(r) \\
            j(r)
        \end{split} \bigg),
\end{align}
where it satisfy the following Dirac equation
\begin{align}
    E_{\mathrm{level}} \Phi_{\mathrm{level}}(r) = 
    \bigg(
        \begin{split}
           M \cos P(r) \qquad  -\partial_r - 2/r + M \sin P(r)   \\
         \partial_r + M \sin P(r) \qquad \qquad\quad \quad- M \cos P(r)
        \end{split} \bigg) \Phi_{\mathrm{level}}(r)
\end{align}

\begin{figure}[htbp]
    \centering
    \includegraphics[width=7.5cm]{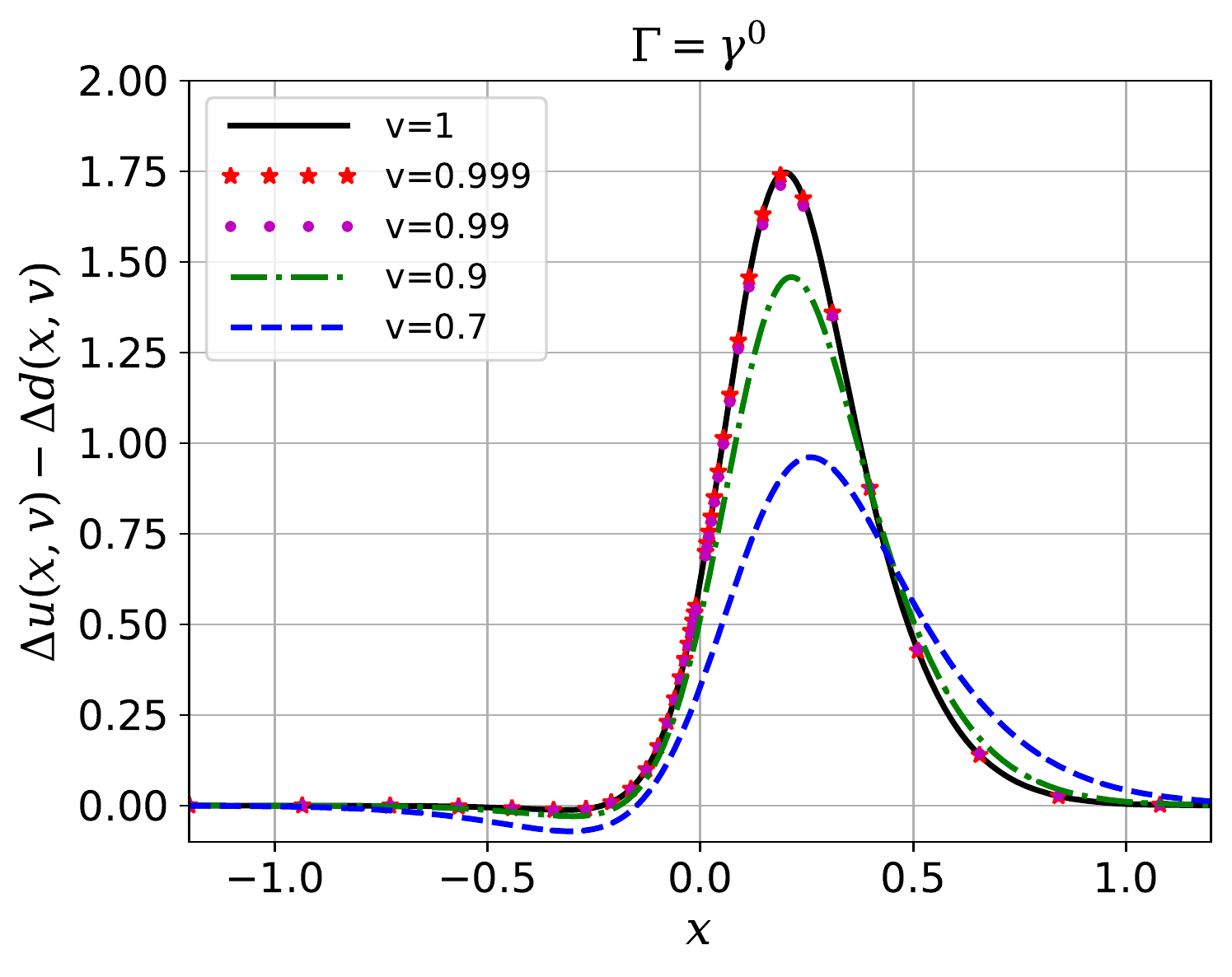}
    \includegraphics[width=7.5cm]{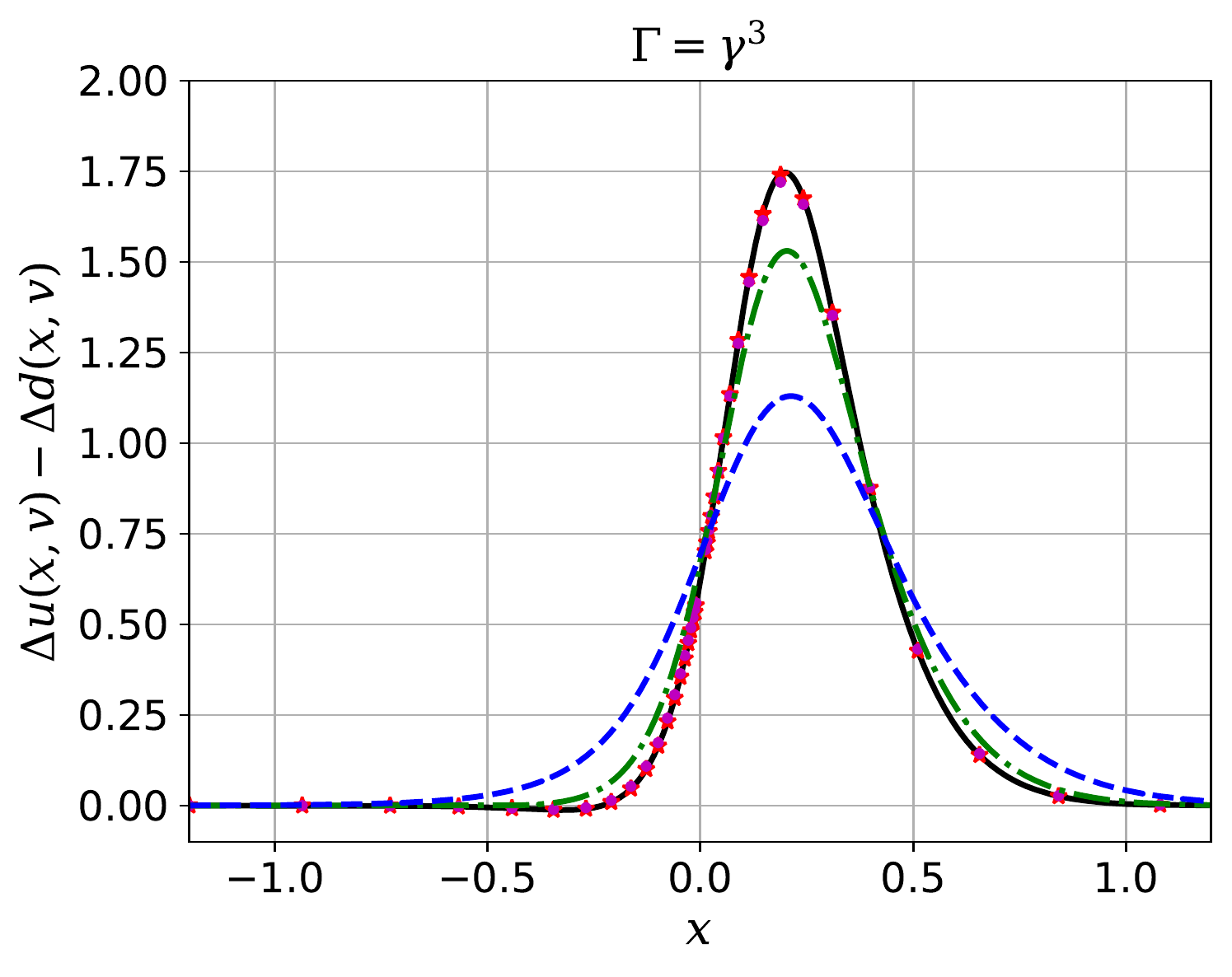}
    \includegraphics[width=7.5cm]{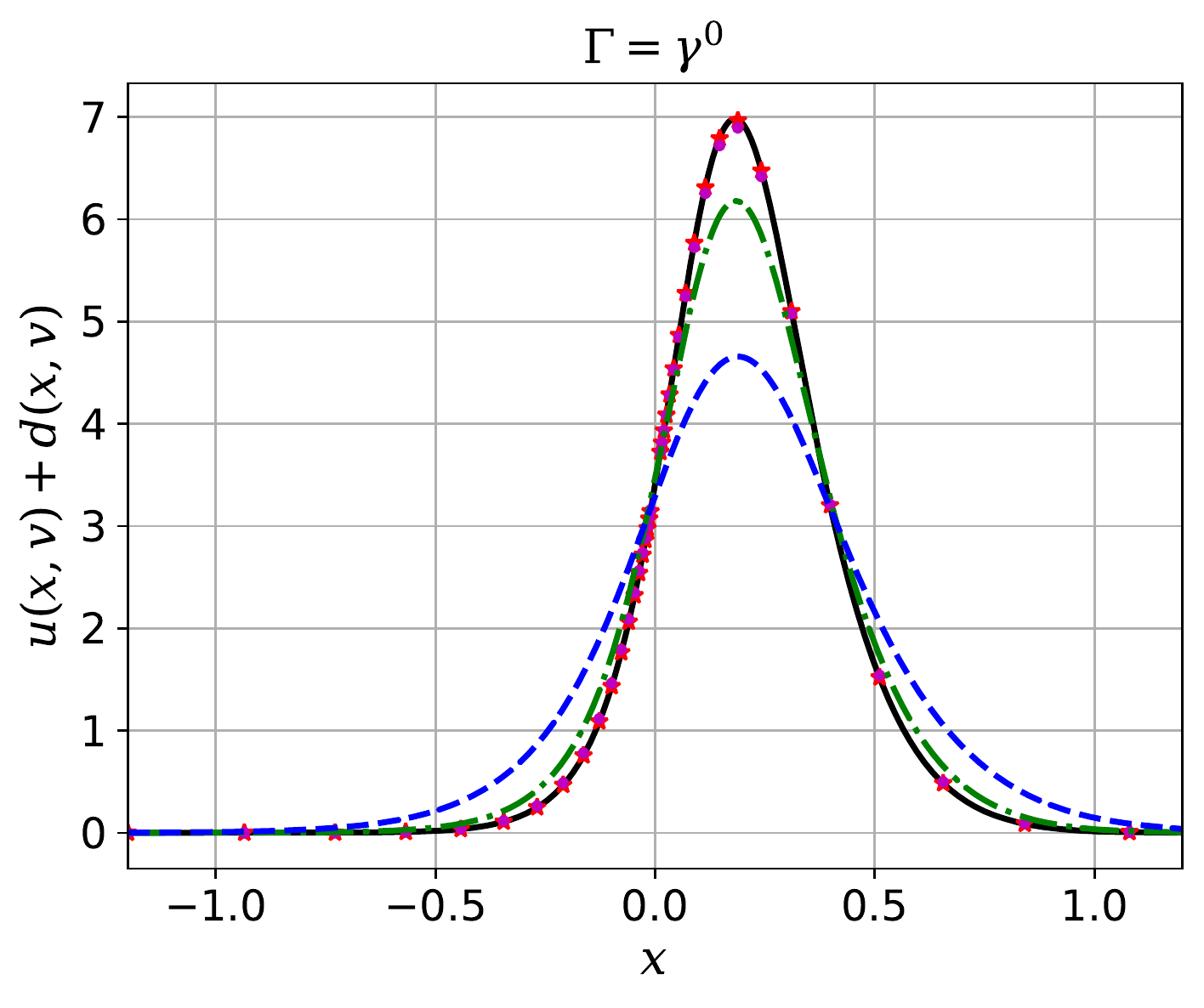}
    \includegraphics[width=7.5cm]{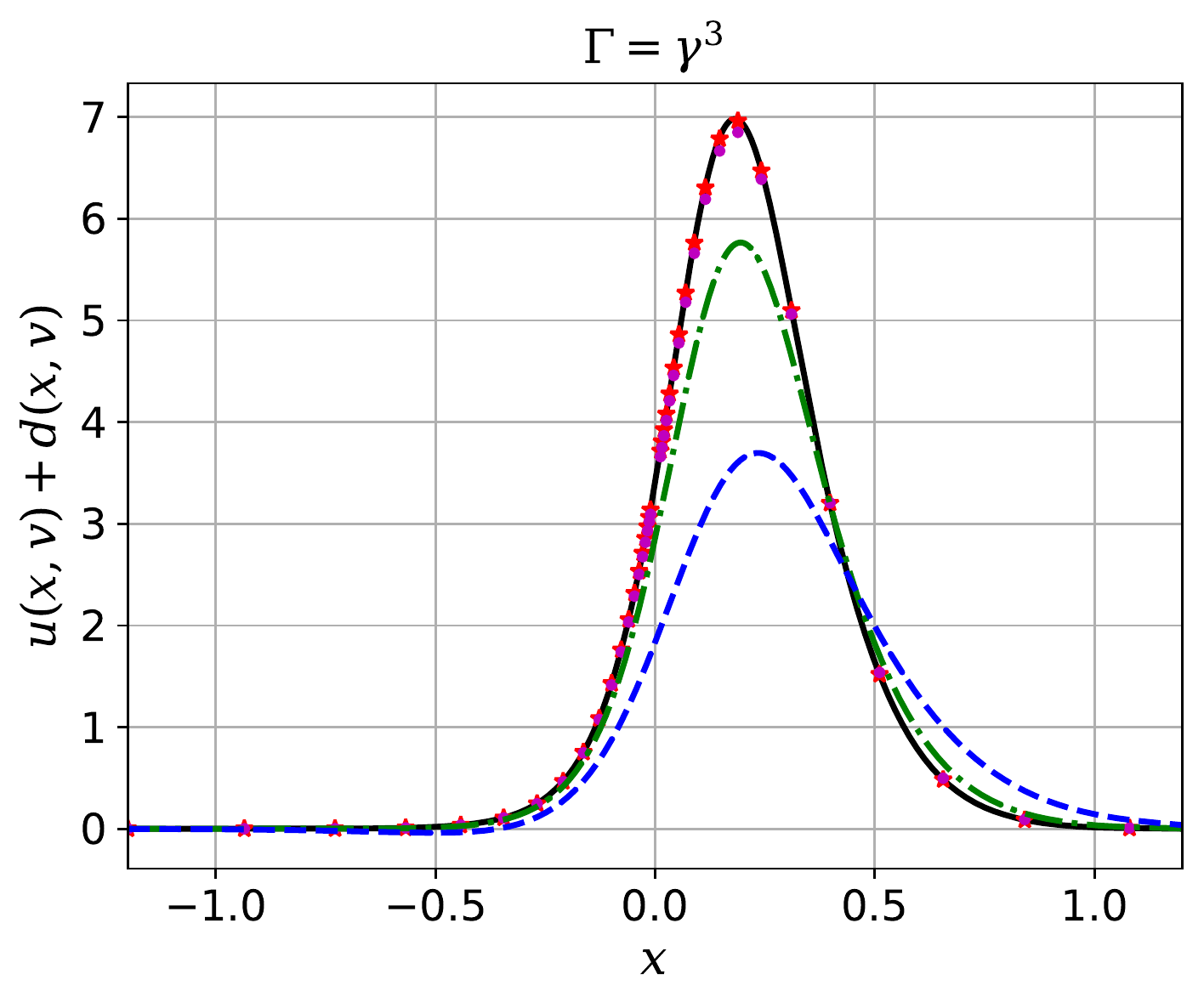}
    \caption{Level contributions to the isovector polarized quasi-PDF (upper) and the isoscalar unpolarized quasi-PDF (lower).}
    \label{fig_appendix_1}
\end{figure}

It is useful to define the following Fourier transformations,
\begin{align}
    h(k) &= \int^\infty_0 dr r^2 h(r) \sqrt{\frac{k}{r}} J_{\half}(kr), \\
    j(k) &= \int^\infty_0 dr r^2 j(r) \sqrt{\frac{k}{r}} J_{1+\half}(kr).
\end{align}
The level contributions to the isoscalar unpolarized (ISU) quasi-PDF are written as \cite{Son:2019ghf}
\begin{align}
   & u(x,v)+d(x,v)_\mathrm{level}  \cr
   &= N_c M_N v \int^\infty_{v|xM_N-E_{\mathrm{level}}|}
    \frac{dk}{2k}\left[h^2(k)+j^2(k)-2 v^2 \frac{xM_N-E_{\mathrm{level}}}{k}
     h(k)j(k) \right] \;(\Gamma=\gamma^0), \\
   &=  N_c M_N v \int^\infty_{v|xM_N-E_{\mathrm{level}}|}
    \frac{dk}{2k}\left[v(h^2(k)+j^2(k))-2 v \frac{xM_N-E_{\mathrm{level}}}{k}
     h(k)j(k) \right] \;(\Gamma=\gamma^3).
\end{align}
The level contributions to the isovector polarized (IVP) quasi-PDF are 
\begin{align}
    & \Delta u(x,v) - \Delta d(x,v)_\mathrm{level}  \cr
    &=\frac{1}{3}
     N_c M_N v^2 \int^\infty_{v|xM_N-E_{\mathrm{level}}|}
     \frac{dk}{2k}\left[h^2(k)+2j^2(k) \frac{v^2(xM_N-E_{\mathrm{level}})^2}{k^2}
     +2 \frac{xM_N-E_{\mathrm{level}}}{k}
      h(k)j(k) \right] \;(\Gamma=\gamma^0), \cr \\
    &=\frac{1}{3}
    N_c M_N v \int^\infty_{v|xM_N-E_{\mathrm{level}}|}
    \frac{dk}{2k}\left[h^2(k)+2j^2(k) \frac{v^2(xM_N-E_{\mathrm{level}})^2}{k^2}
    +2v^2 \frac{xM_N-E_{\mathrm{level}}}{k}
     h(k)j(k) \right] \;(\Gamma=\gamma^0). \cr
 \end{align}
In Fig. \ref{fig_appendix_1}, the level contributions to the IVP (upper) and ISU (lower) quasi distributions are depicted for the Dirac matrix $\Gamma=\gamma^0$ (left) and $\Gamma=\gamma^3$ (right). 
\subsection{Continuum contributions}
\begin{figure}[htbp]
    \centering
    \includegraphics[width=7.5cm]{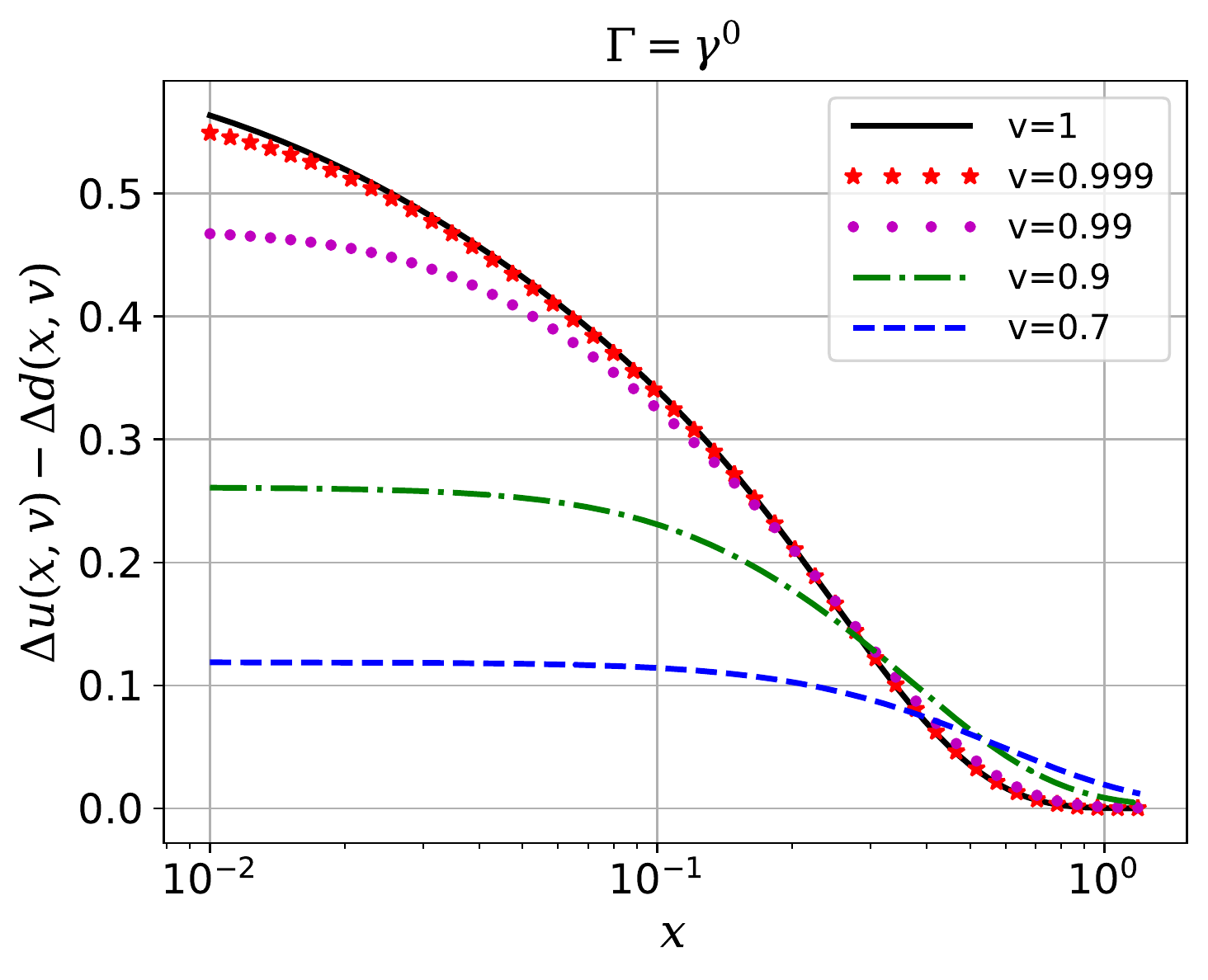}
    \includegraphics[width=7.5cm]{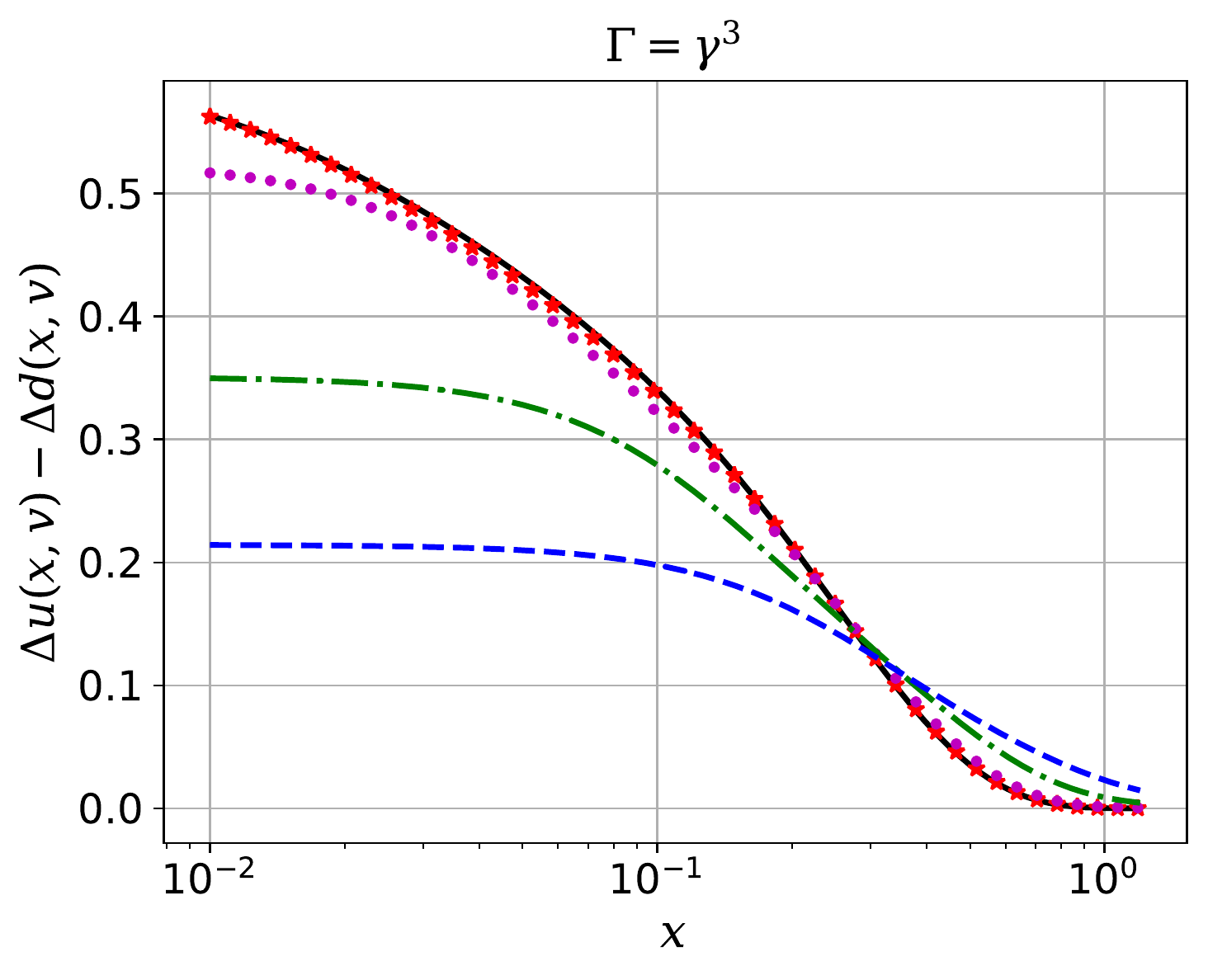}
    \includegraphics[width=7.5cm]{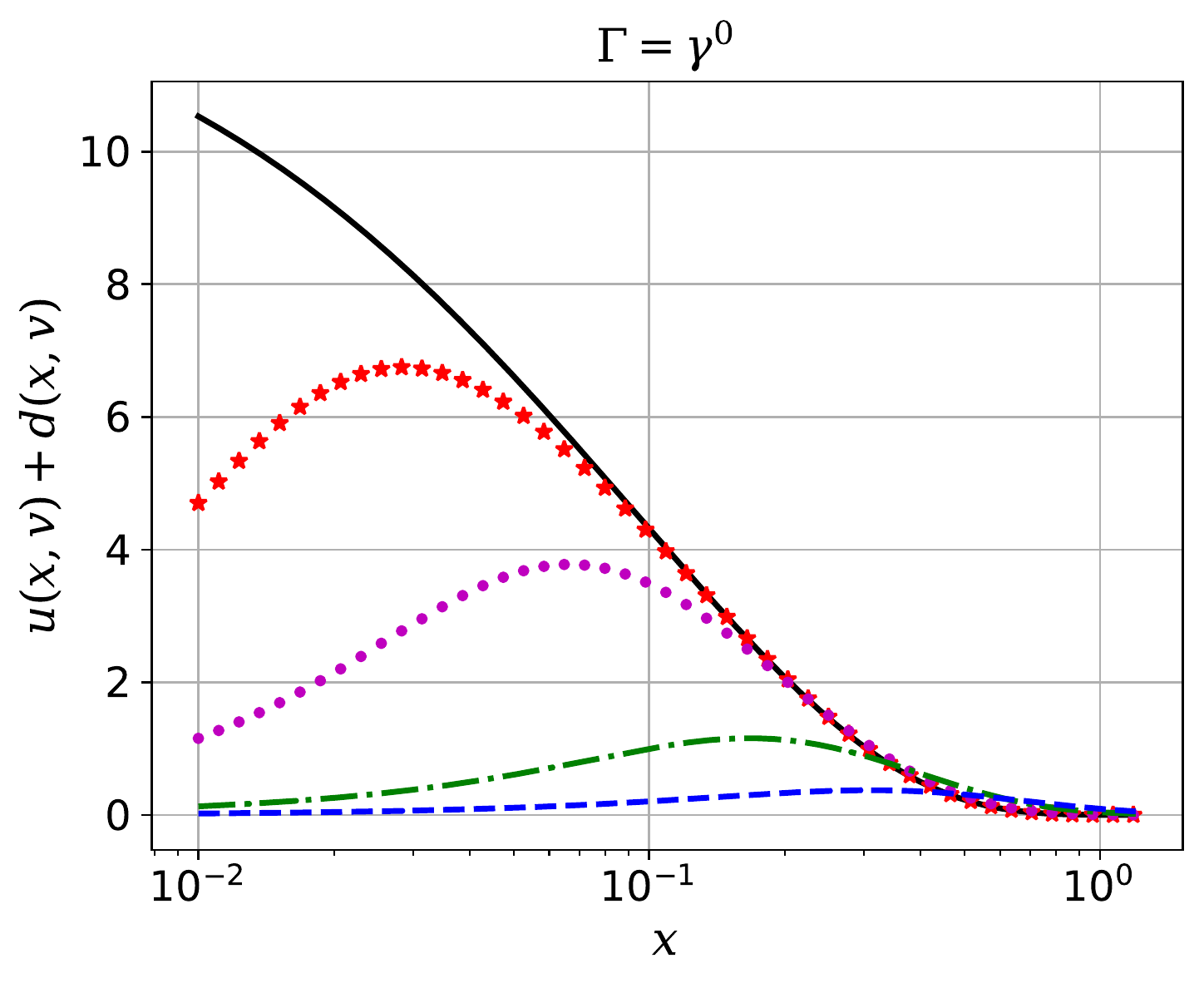}
    \includegraphics[width=7.5cm]{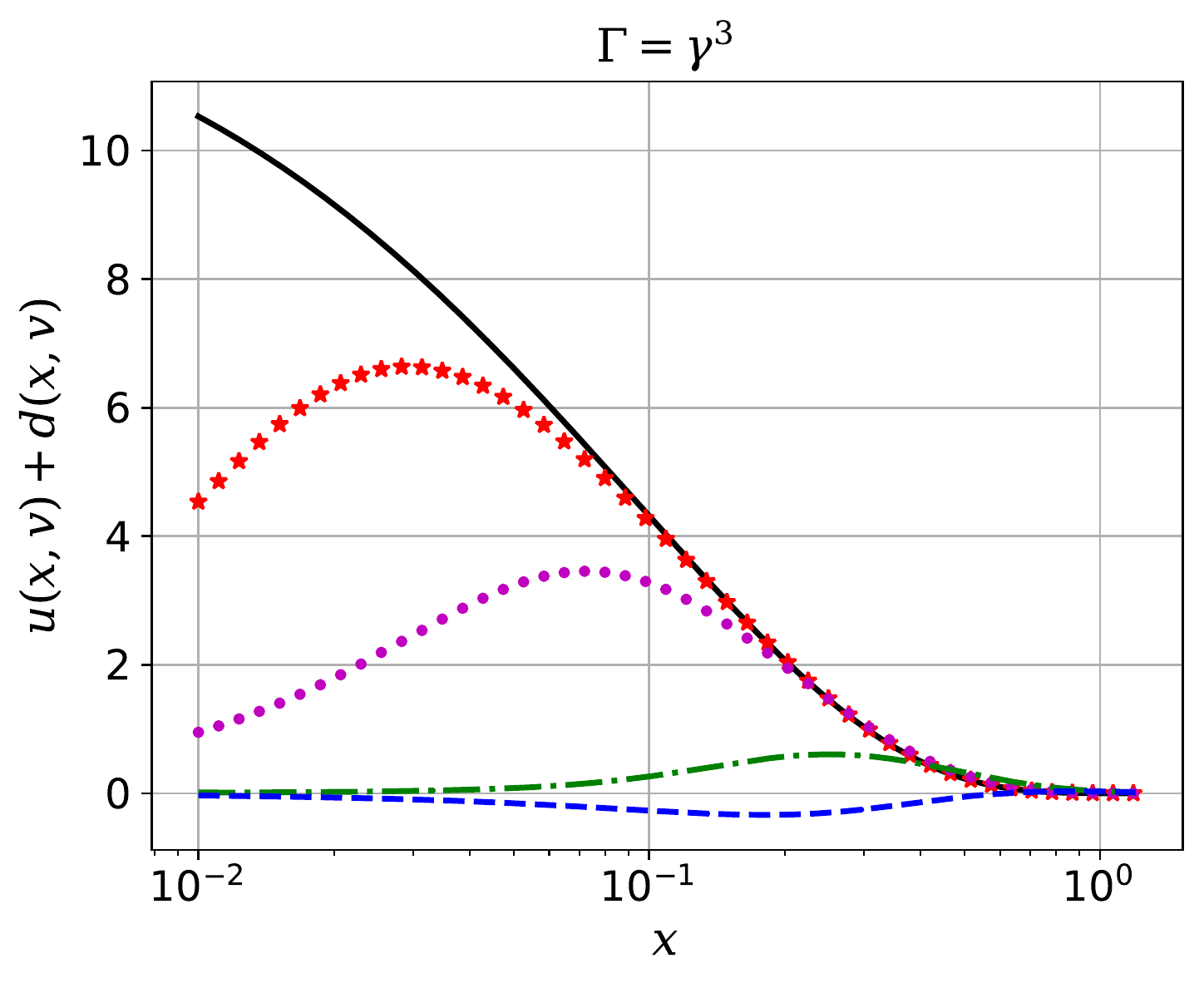}
    \caption{Continuum contributions to the isovector polarized quasi-PDF (upper) and the isoscalar unpolarized quasi-PDF (lower).}
    \label{fig_appendix_2}
\end{figure}

For completeness, we plot in Fig. \ref{fig_appendix_2} the continuum contributions to the IVP (upper) and ISU (lower) quasi distributions for the Dirac matrix $\Gamma=\gamma^0$ (left) and $\Gamma=\gamma^3$ (right). One observes that the behavior of the quasi-PDFs at small $x$ region in the nucleon boost velocity $v$ mostly depends on the continuum contributions. 

\bibliography{ivp_manuscript_plb}

\end{document}